# Diode effect in a medium with a helical magnetic structure

Andrey Alexandrovich Fraerman and Oleg Georgievich Udalov
Institute for physics of microstructures RAS, GSP-105, Nizhny Novgorod, 603950
Russia

The volt-ampere characteristics of a medium with helical magnetic structure are theoretically studied. Phenomenological and microscopic theories of a diode effect in such a medium are proposed. Diode effect appears only in the medium with noncoplanar magnetic structure. Two mechanisms for the diode effect are considered. They are electron energy spectrum asymmetry and electron scattering asymmetry.

72.15.Gd, 75.47.-m

The interplay between the spin and spatial degrees of freedom of electrons is of great interest from both a theoretical and a practical point of view [1]. Investigations into this interplay are based on two well-known effects: spin-orbit coupling and Zeeman splitting. In the presence of spin-orbit coupling, the energy spectrum of an electron $E_\sigma(\vec{p})$ in a crystal obeys the relation

$$E_\sigma(\vec{p}) = E_{\sigma'}(-\vec{p}), \tag{1}$$

where $\vec{p}$ is the quasi-momentum and $\sigma$ and $\sigma'$ are the different spin states of the electron. When the Hamiltonian contains the Zeeman term (for example, for ferromagnetic metals in the frame of the s-d approximation), the symmetry of an electron energy spectrum differs from Eq. (1) and depends on the spatial distribution of a magnetic field (or magnetization). If the magnetic field is coplanar, one can transform Hamiltonian of the system to the real form by using the corresponding rotation of a spin coordinate system. Consequently, the electron spectrum has the symmetry

$$E_\sigma(\vec{p}) = E_\sigma(-\vec{p}). \tag{2}$$

Note that in contrast to Eq. (1), Eq. (2) implies that each branch of the spectrum is an even function of $\vec{k}$, but the spin degeneracy is removed. For a non-coplanar magnetic field, Eqs. (1) and (2) are not valid and the electron spectrum is a general function of quasi-momentum. Thus, momentum-reversal symmetry is broken in a non-coplanar system. This can lead to peculiarities in the transport properties of crystals in a non-coplanar magnetic field (or with a non-coplanar magnetization distribution).

Little research has been conducted on the consequences of spectrum asymmetry with respect to inverting the direction of current flow. In [2, 3], a mesoscopic metal ring with a magnetic texture was studied. These studies found that when the distribution of magnetization in the ring becomes non-coplanar, persistent electrical current occurs in the ring. The magnitude of this current is proportional to the degree of non-complanarity of the magnetization distribution. This effect originates from spectrum quantization in the ring. The magnitude of the persistent current approaches zero as the ring size increases (even in the absence of inelastic scattering).

The aim of this letter is to investigate the possible consequences of momentum-reversal symmetry-breaking in macroscopic systems with a non-coplanar magnetic field distribution. Specifically, we investigate a diode effect in a medium with a helical magnetic structure. In the first part of this letter, the general phenomenological theory of a diode effect in a medium with an arbitrary magnetization distribution will be considered. In the second part, we propose a simple microscopic theory of the diode effect in a medium with a helical magnetic structure.

Consider a medium with a periodic magnetization distribution $\vec{M}(\vec{r})$. The relationship determining the current component $\vec{j}^{(2)}$, which is quadratic in electric field ($\vec{E}$), is

$$j_i^{(2)} = \gamma_{ijk} E_j E_k,  \qquad (3)$$

where $\gamma_{ijk}$ are components of the third-rank conductivity tensor. Inverting the coordinates reverses the current direction. Therefore $\gamma_{ijk}$ should contain an odd number of spatial derivatives (note that $\vec{M}(\vec{r})$ is an axial vector). Let us suppose that the only force in the system is exchange interaction and that there are no relativistic effects. Exchange forces depend only on the mutual orientation of spins; therefore, simultaneous rotation of all spins in the system by the same angle does not change any physical characteristics of the system [4]. This means that in the expressions for the current magnetic and spatial vector indexes should not be convolved with one another [5]. The restrictions determine the tensor structure unambiguously

$$\gamma_{ijk} = \alpha\left(\vec{M}, \left[\frac{\partial \vec{M}}{\partial x_i} \times \frac{\partial^2 \vec{M}}{\partial x_j \partial x_k}\right]\right), \qquad (4)$$

here $\alpha$ is constant and does not depend on the magnetization $\vec{M}(\vec{r})$. Note that the restrictions mentioned above do not forbid tensors containing the magnetization vector $\vec{M}(\vec{r})$ twice, but for a magnetization distribution that is periodic in space, such a tensor will give a period-average current $\vec{j}^{(2)}$ of zero.

It is obvious that the conductivity tensor $\gamma$ is non-zero only in the case of a non-coplanar magnetization distribution. Thus, in systems with exchange interaction and a non-coplanar magnetization distribution, the diode effect can exist.

Let us apply Eq. (4) to a material with helical magnetic structure. The helical magnetization distribution is described by the formula

$$\vec{M} = \left(\sqrt{1-m^2}\cos(qz), \sqrt{1-m^2}\sin(qz), m\right). \qquad (5)$$

Combining Eq. (5) into Eq. (4, 3) one easily gets

$$j_z^{(2)} = \alpha m(1-m^2)q^3 E^2, \quad j_{x,y}^{(2)} = 0. \qquad (6)$$

Now let us demonstrate mechanisms of diode effect arising in a medium with a helical magnetic structure from the microscopic point of view.

Let us first examine the energy spectrum of electrons in a medium with a helical magnetic structure. In the frame of the s-d model, the conduction electron behavior is described by the equation

$$-\Delta \Psi(\vec{r}) - \widetilde{J}\vec{M}(z)\hat{\sigma}\Psi(\vec{r}) = \widetilde{\varepsilon}\,\Psi(\vec{r}), \tag{7}$$

where $\vec{M}(z)$ is a unit vector directed along the local magnetic moment, $\hat{\sigma}$ is the Pauli vector, $\widetilde{J} = J/(\hbar^2/2m_e)$, where $J$ is a constant of s-d exchange, and $\widetilde{\varepsilon} = \varepsilon/(\hbar^2/2m_e)$, where $m_e$ is effective electron mass.

From Eq. (7), it is easily shown that, in the case of a coplanar magnetization distribution ($m = 0,1$ or $q = 0$), it is always possible to find a system of spin coordinates where Eq. (7) is invariant with respect to the complex conjugation. In this case, the electron spectrum is an even function with a quasi-momentum. Note that the value of $qm(1-m^2)$ can be given as the degree of non-coplanarity. If the magnetization distribution is non-coplanar, Eq. (7) is not invariant with respect to the complex conjugation. This means that the restriction on the evenness of the spectrum is not valid in such a system. As shown in [6, 7] the wave function of Eq. (7) is planar:

$$\Psi_{\vec{k}}^{\pm}(\vec{r}) = \frac{1}{\sqrt{1+(\delta^{\pm}(k_z))^2}} \begin{pmatrix} \delta^{\pm}(k_z) e^{-i(q/2)z} \\ e^{i(q/2)z} \end{pmatrix} e^{i\vec{k}\vec{r}}, \tag{8}$$

$$\delta^{\pm}(k) = \frac{m\widetilde{J} - qk_z \pm \sqrt{q^2 k_z^2 + \widetilde{J}^2 - 2\widetilde{J}mqk_z}}{\widetilde{J}(1-m^2)^{1/2}}. \tag{9}$$

The electron state is determined by a wave vector $\vec{k}$ and by a spin state ($^+$ or $^-$). The spin of the particle rotates with the magnetization. It is easy to show that the angle between a magnetization vector and a particle spin $\Delta\phi$ at any point can be calculated by the formula $\cos(\Delta\phi) = \frac{1}{\delta^2+1}[m(\delta^2-1) - 2\delta\sqrt{1-m^2}]$. Angle $\Delta\phi \neq 0$ and depends on $k_z$ and parameters of the magnetic structure. If $q \to 0$ or $k_z \to \infty$ (the case of the magnetization which varies slowly in the space), then $\Delta\phi \to 0$ for the $^-$ spin state (a lower branch of the energy spectrum) and $\Delta\phi \to \pi$ for the $^+$ spin states (an upper branch of the energy spectrum).

The energy spectrum of electrons $\widetilde{\varepsilon}^{\pm}(\vec{k})$ is determined by the expression

$$\widetilde{\varepsilon}^{\pm}(\vec{k}) = k_x^2 + k_y^2 + \widetilde{\varepsilon}^{\pm}{}_z(k_z) = k_x^2 + k_y^2 + k_z^2 + \frac{q^2}{4} \pm \sqrt{q^2 k_z^2 + \widetilde{J}^2 - 2\widetilde{J}mqk_z}. \tag{10}$$

Eq. (10) shows that under the condition of $qm \neq 0$, the spectrum is an asymmetric function of quantum number $k_z$. Beforehand note that in the work a case of slowly rotating magnetic field is considered

$$\beta = \frac{qk_z}{\tilde{J}} << 1. \tag{11}$$

Under the condition (11) electron spectrum can be expanded in the series

$$\varepsilon^{\pm}(\vec{k}) \approx k_x^2 + k_y^2 + q^2/4 \pm \tilde{J} \mp qmk_z + (1 \pm 0.5q^2/\tilde{J})k_z^2 \pm \eta k_z^3 \tag{12}$$

$$\eta = \frac{1}{2}mq\frac{q^2}{\tilde{J}^2} \tag{13}$$

First three terms in this expression give us simple parabolic spectrum, which shifted to the left and to the right for different spectrum branches. Last term of (12) provides spectrum asymmetry cubical in $\beta$.

Second point that we examine here is scattering of electrons in magnetic spiral by potential (nonmagnetic) impurities and phonons. In the frame of Born approximation expressions for transition probabilities $W(\vec{k}_1, \vec{k}_2)$ have the form

$$W_{++}(\vec{k}_1,\vec{k}_2) = |A_{++}|^2 w_{k_1,k_2}, \quad W_{--}(\vec{k}_1,\vec{k}_2) = |A_{--}|^2 w_{k_1,k_2}, \quad W_{+-}(\vec{k}_1,\vec{k}_2) = |A_{+-}|^2 w_{k_1,k_2}, \tag{14}$$

$$A_{++}(\vec{k}_1,\vec{k}_2) = \frac{\delta^+(k_{1z})\delta^+(k_{2z}) + 1}{\sqrt{1+(\delta^+(k_{1z}))^2}\sqrt{1+(\delta^+(k_{2z}))^2}} =$$
$$= 1 - 1/8(\beta_1 - \beta_2)^2 + m\beta_1\beta_2(\beta_1 + \beta_2)9/32 - m(\beta_1^3 + \beta_2^3)5/16 = A_{--}(\vec{k}_1,\vec{k}_2) \tag{15}$$

$$A_{+-}(\vec{k}_1,\vec{k}_2) = \frac{\delta^+(k_{1z})\delta^-(k_{2z}) + 1}{\sqrt{1+(\delta^+(k_{1z}))^2}\sqrt{1+(\delta^-(k_{2z}))^2}} =$$
$$= 1/2(\beta_1 - \beta_2)(1 + m(\beta_1 + \beta_2)) = A_{-+}(\vec{k}_2,\vec{k}_1) \tag{16}$$

$w_{k_1,k_2}$ are usual transition probabilities for electron-impurities and electron-phonon scattering [8], $\beta_{1,2} = \frac{qk_{1z,2z}}{\tilde{J}}$. Lets note, that $w_{k_1,k_2}$ is symmetrical function i.e. $w_{-k_{1z},-k_{2z}} = w_{k_{1z},k_{2z}}$. Nontrivial part $A(\vec{k}_1,\vec{k}_2)$ of scattering probability arises from complicate structure of electron wave function in a magnetic helix. This factor leads to anisotropy of scattering (z-axis differs from x- and y-axis). But more important is that this factor is asymmetrical function of $k_z$ ($W(-k_{1z},-k_{2z}) \neq W(k_{1z},k_{2z})$). It is easy to see that scattering asymmetry arises due to absence of complex conjugation invariance of Schrodinger equation for electron in noncoplanar magnetic field and disappears in coplanar system ($mq=0$). Note that scattering asymmetry is also proportional to $m\beta^3$.

Thus, absence of invariance of Schrodinger equation with respect to complex conjugation for electron in noncoplanar magnetic field leads to electron spectrum and electron scattering asymmetries.

Now we calculate the current flowing in the medium with a helical magnetic structure when an external electric field $E$ is applied along the helix axis $z$.

Lets assume that description of electron transport in frame of Boltzman equations are valid [9], then

$$-eE\frac{\partial f^{\pm}(\vec{k})}{\partial \hbar k_z} = \frac{V}{8\pi^3}\int d^3k_2 W_{\pm\pm}(\vec{k}_1,\vec{k}_2)(f^{\pm}(\vec{k}_1)-f^{\pm}(\vec{k}_2)) + \\ + \frac{V}{8\pi^3}\int d^3k_2 W_{\pm\mp}(\vec{k}_1,\vec{k}_2)(f^{\pm}(\vec{k}_1)-f^{\mp}(\vec{k}_2))$$, (17)

where $f^{\pm}(\vec{k})$ are distribution functions of electrons, $\pm$ represents different spin states. $V$ is volume of the system. First term of collision integral describes electron transitions on the same spectrum branch. Second term describes transitions from one spectrum branch to another. Transition probability $W(\vec{k}_1,\vec{k}_2)$ includes both elastic $W_{el}(\vec{k}_1,\vec{k}_2)$ and inelastic $W_{in}(\vec{k}_1,\vec{k}_2)$ scattering. Lets assume that there are only interaction of electrons with phonons (inelastic processes providing energy relaxation and appearing only in the second order in electric field) and potential (not magnetic) impurities (which provides elastic scattering and momentum relaxation). Note that in the Born approximation one always has $W(\vec{k}_1,\vec{k}_2) = W(\vec{k}_2,\vec{k}_1)$. In (17) it is assumed that temperature of our system $T \gg T_D$ ($T_D$ is Debay temperature).

Expression for electric current has usual form

$$j = e\frac{1}{8\pi^3}\frac{\hbar}{2m_e}\int f^+(\vec{k})\frac{\partial \tilde{\varepsilon}^+(k_z)}{\partial k_z}d^3k + e\frac{1}{8\pi^3}\frac{\hbar}{2m_e}\int f^-(\vec{k})\frac{\partial \tilde{\varepsilon}^-(k_z)}{\partial k_z}d^3k \quad (18)$$

Assuming that electrical field $E$ is sufficiently small, we have the series for distribution function of electrons

$$f^{\pm}(\vec{k}) = f^0(\vec{k}) + eEf^{\pm(1)}(\vec{k}) + (eE)^2 f^{\pm(2)}(\vec{k}) + ... \quad (19)$$

Substituting expansions (19) and (12) into (18) the following expression for diode term can be obtained

$$j = j^{(1)+} + j^{(2)+} + j^{(1)-} + j^{(2)-} + ...$$

$$j^{(2)\pm} = e(eE)^2 \frac{1}{8\pi^3}\frac{\hbar}{2m_e}\left\{\int (k_z \mp k_0)f^{\pm(2)}(\vec{k})d^3k \pm 3\alpha\int k_z^2 f^{\pm(2)}(\vec{k})d^3k\right\}, \quad (20)$$

here $k_0 = \frac{qm}{2(1+q^2/2\tilde{J})}$. First term of (20) appears from parabolic part of electron spectrum, but it is not a zero only if $f^{(2)}$ has an odd part as function of $k_z$. Second term of (20) appears only due to spectrum asymmetry and it is non-zero if $f^{(2)}$ has an even part as function of $k_z$.

Estimation of even and odd part of $f^{(2)}$ can be made on the base of Boltzman equation (17) using approximation of quasielastic scattering [10,11]. In the frame of this approximation it is supposed that momentum relaxation is much faster than energy relaxation. For evaluation of even part of $f^{(2)}$ one can neglect spectrum and scattering asymmetries. In the approximation of quasielastic scattering one has

$$f_{even}^{\pm(2)}(\vec{k}) = c_1(eE)^2 \tau^2 \left(\frac{\hbar k_z}{m_e}\right)^2 \frac{\partial^2 f_0(\varepsilon^\pm)}{\partial \varepsilon^{\pm 2}}, \text{ where} \qquad (21)$$

$$\tau = \int W_{el}(\vec{k}_1, \vec{k}_2) d^3 k_2 \qquad (22)$$

here $f^0$ is Fermi distribution. Odd part of $f^{(2)}$ can be evaluated as a perturbation of $f_{even}^{\pm(2)}(\vec{k})$ due to small scattering asymmetry

$$f_{odd}^{\pm(2)}(\vec{k}) = \frac{(eE\tau)^2}{m_e} \frac{mq^3}{\tilde{J}^3} \left\{ c_2 k_z^3 \frac{\partial f_0(\varepsilon^\pm)}{\partial \varepsilon^\pm} + \frac{\hbar^2}{m_e} k_z \frac{\partial^2 f_0(\varepsilon^\pm)}{\partial \varepsilon^{\pm 2}} \left( c_3 |\vec{k}|^4 + c_4 k_z^2 |\vec{k}|^2 + c_5 k_z^4 \right) \right\}$$

(23)

$c_i$ are numerical coefficients whose are not important in our qualitative consideration. Substituting (21) and (23) into (20) one can calculate quadratic in the electric field part of current

$$j^{(2)} = e \frac{(eE\tau)^2}{m_e \hbar} m \frac{q^3 k_f^3}{\tilde{J}^3} (\chi_1 + \chi_2 \frac{J^2}{\varepsilon_f^2}) k_f^2 =$$

$$= \sigma_L \frac{p_E}{p_f} m \beta_f^3 (\chi_1 + \chi_2 \frac{J^2}{\varepsilon_f^2}) E \qquad (24)$$

here $\sigma_L$ is linear conductivity, $p_E = eE\tau$, $\beta_f = qk_f/\tilde{J}$, $\chi_i$ are numerical coefficients. First term of nonlinear current arises from scattering asymmetry and it is essentially bigger than second term arising from spectrum asymmetry (group velocity asymmetry).

It is seen from (24) that expression for quadratic in the electric field part of a current derived from our simple microscopic model has the same form as the expression obtained in phenomenological theory.

Thus, in the medium with noncoplanar magnetization distribution the diode effect appears. Microscopic reasons for it are electron spectrum and scattering asymmetries.

Note that helical magnetic structure realizes in a number of materials. Among them rare-earth elements (Ho, Dy, Er [7,12]) and some alloys (MnSi [13]).

In this letter, we describe the mechanism of a diode effect occurring in media containing helical magnetic structures. A phenomenological theory of a diode effect in magnetic media was first considered. When the magnetization distribution is non-coplanar, the volt-ampere characteristics of the medium may contain a term proportional to the square of the electric field. A simple microscopic model of a diode effect in a medium with a helical magnetic structure was proposed. This model confirms the result of the phenomenological theory. The microscopic reasons for a diode effect are momentum-reversal symmetry breaking leading to group velocity asymmetry and electron scattering asymmetry.

We acknowledge useful discussion with V.Ya. Aleshkin, M.V. Sapozhnikov. The work was supported by INTAS 03-51-4778 and RFBR 07-02-01321-a

[1] I. Zutic, J. Fabian, S. D. Sarma, Rev. Mod. Phys. **76**, 323 (2004)
[2] D. Loss, P. Goldbart, A.V. Balatsky, Phys. Rev. Lettt. **65**, 1655 (1990)
[3] G. Tatara, H. Kohno, Phys. Rev. B, **67**, 113316 (2003)
[4] A.F. Andreev, V.I. Marchenko, Sov Phys Uspekhi, **23** (1), 21 (1980)
[5] L.D. Landau, E.M. Lifshitz, *Electrodynamics of Continuous Media* (v. 8 of *Course of Theoretical Physics*)(2ed., Pergamon, 1984)
[6] M. Calvo, Phys. Rev. B **19**, 5507 (1978)
[7] E. L. Nagaev, *Physics of Magnetic Semiconductors* (Mir, Moscow, 1983)
[8] J.M. Ziman, *Principles of the Theory of Solids* (Cambridge, 1972)
[9] W. Kohn, J.M. Luttinger, Phys. Rev., **108**, 3, 590 (1957)

[10] V.L. Ginzburg, A V Gurevich, *SOV PHYS USPEKHI*, 1960, **3** (2), 175-194.

[11] F.G. Bass, Yu.G. Gurevich, *SOV PHYS USPEKHI*, 1971, **14** (2), 113-124.
[12] W.C. Koehler, J.W. Cable, M.K. Wilkinson and E.O. Wollan, Phys. Rev. **151**, 414 (1966)
[13] Y. Ishikawa, G. Shirane, J. A. Tarvin et al., Phys. Rev. B **16**, 4956 (1977)